\documentclass[iop]{emulateapj}
\usepackage{amsmath}
\usepackage{inputenc}
\usepackage{array}   
\usepackage{hhline}
\usepackage{color}
\usepackage[sort&compress]{natbib}
\usepackage{graphicx}

\newcommand{\mbfB}{\mathbf{B}}
\newcommand{\mbfD}{\mathbf{D}}
\newcommand{\mbfE}{\mathbf{E}}
\newcommand{\mbfF}{\mathbf{F}}
\newcommand{\mbfJ}{\mathbf{J}}

\newcommand{\mbfvv}{\mathbf{v}}

\newcommand{\mbfk}{\mathbf{k}}

\newcommand{\mbfp}{\mathbf{p}}
\newcommand{\mbfvA}{\mathbf{v}_A}

\newcommand{\mbfnabla}{\mathbf{\nabla}}

\newcommand{\eps}{\epsilon}

\usepackage{tabularx}
\usepackage{graphicx}
\usepackage[breaklinks,colorlinks,citecolor=blue,linkcolor=magenta]{hyperref}

\shorttitle{Pressure Anisotropy in Cosmic Ray Hydrodynamics}
\shortauthors{Zweibel}

\begin{document}

\title{The Role of Pressure Anisotropy in Cosmic Ray Hydrodynamics}

\author{Ellen G. Zweibel\altaffilmark{1,2}}

\altaffiltext{1}{Department of Astronomy, U Wisconsin-Madison, 475 N Charter St., Madison,
                 WI 53706, U.S.A.}
\altaffiltext{2}{Department of Physics, U. Wisconsin-Madison, 1150 University Ave., Madison, WI 53706 U.S.A.}

\begin{abstract}
Cosmic ray propagation in the Milky Way and other galaxies is largely diffusive, with mean free path determined primarily by pitch angle scattering from hydromagnetic waves with wavelength of order the cosmic ray gyroradius.  In the theory of cosmic ray self confinement, the waves are generated by instabilities driven by the cosmic rays themselves. The dominant instability is due to bulk motion, or streaming, of the cosmic rays, parallel to the background magnetic field $\mbfB$, and transfers cosmic ray momentum and energy to the thermal gas as well as confining the cosmic rays. Classical arguments and recent numerical simulations show that self confinement due to the streaming instability breaks down unless the cosmic ray pressure and thermal gas density gradients parallel to $\mbfB$ are aligned, a condition which is unlikely to always be satisfied We investigate an alternative mechanism for cosmic ray self confinement and heating of thermal gas based on pressure anisotropy instability. Although pressure anisotropy is demonstrably less effective than streaming instability as a self confinement and heating mechanism on global scales, it may be important on mesoscales, particularly near sites of cosmic ray injection.
\end{abstract}


\section{Introduction}\label{s:introduction}

Cosmic rays  provide a window on high energy processes throughout the Universe, significantly affect interstellar and intracluster gas dynamics and energy balance,  and are agents of star formation and black hole feedback. All these aspects of cosmic ray astrophysics depend on how cosmic rays propagate through the ambient magnetic field $\mbfB$. 

The near isotropy and long confinement times of Galactic cosmic rays imply that their propagation is largely diffusive (see \cite{grenier15} for a recent review). While diffusion through space can be produced by propagation along randomly wandering magnetic field lines $\mbfB$,  diffusion parallel to $\mbfB$, and to a lesser extent perpendicular to $\mbfB$, is thought to be primarily due to scattering by magnetic fluctuations on scales of order the cosmic ray gyroradius. 

There are two theories for the origin of these fluctuations. In  the self confinement theory, the fluctuations are hydromagnetic waves that have been amplified by kinetic instabilities driven by cosmic ray momentum space anisotropy (\cite{wentzel68,kulsrudpearce69}).  The unstable feature, or free energy source for the instability, is generally  bulk drift, or streaming, which arises naturally, e.g, from the presence of discrete cosmic ray sources and global Galactic gradients. In  the extrinsic turbulence theory, the fluctuations are also hydromagnetic waves, but are driven by a mechanism such as a turbulent cascade that is independent of cosmic
rays. 

In both theories, momentum is transferred between the cosmic rays and the thermal gas through what can be described in the limit of short scattering mean free path as a pressure gradient
force $-\mbfnabla P_c$. In self confinement theory, the thermal gas is also heated collisionlessly by damping the waves  at the rate the cosmic rays excite them; for the streaming instability, this works out to be $\vert\mbfvA\cdot\mbfnabla P_c\vert$.  In the extrinsic turbulence theory, there is no
heating, provided the fluctuations have no preferred direction of propagation. Here, $P_c$ and $\mbfvA$ are the cosmic ray pressure and Alfven velocity in the plasma component $\mbfB/\sqrt{4\pi\rho_i}$, with
$\rho_i$ the plasma mass density. Both theories lend themselves to fluid descriptions of cosmic ray interactions with thermal gas
(``cosmic ray hydrodynamics"), and can be smoothly bridged  when both extrinsic and self generated waves are present \citep{zweibel17}.
Because both theories are based on scattering, they include spatial diffusion, but in many cases it is weak compared to advection by the
thermal gas or Alfv\'enic streaming relative to it. 

Both theories have been implemented in models of galactic winds and star formation feedback \cite{breitschwerdtetal91,everettetal08,uhligetal12,agertzetal13,boothetal13,salembryan14,girichidisetal16,ruszkowskietal17,farberetal18,maoostriker18,chanetal19}. These works show that the mass flux, momentum flux,  thermal structure, and even the existence of galactic winds are sensitive to the model of cosmic ray transport, as is the degree to which cosmic ray feedback suppresses star formation.  For example, assuming that
cosmic rays are advected with the thermal gas but neither stream nor diffuse suppresses wind launching in Milky Way like disks \cite{uhligetal12}, but lowers the star formation rate more than models with cosmic ray streaming, which are more effective in launching winds
but less effective in suppressing star formation \cite{ruszkowskietal17}.  With observational constraints on cosmic ray transport in star forming galaxies now emerging from models of their $\gamma$-ray emission \cite{chanetal19}, it is more important than ever to understand all the
physical processes in play. 

The streaming instability can be excited when the bulk drift speed $v_D$ is super-Alfvenic. The growth rate increases with $v_D$, and cosmic rays can be considered self confined if the drift required to overcome damping is not too much greater than $v_A$. While instability growth and damping rates depend only on local conditions which can be evaluated from point to point,  self confinement also depends on the global  structure of the system due to a ``bottleneck effect" which was first hypothesized
by \cite{Skilling71} and first demonstrated in numerical simulations by \cite{wieneretal2017}. The drift anisotropy is associated with a spatial gradient in  cosmic ray pressure $P_c$
along the background magnetic field $\mbfB$
such that the drift is down the gradient and the unstable waves propagate in the same direction as the drift. It can then be shown that $P_c$ varies along a magnetic flux tube in proportion to $\rho_i^{\gamma_c/2}$, where $\gamma_c\sim 4/3$ is the  cosmic ray adiabatic index. If $\rho_i$ increases in the direction of cosmic ray streaming,  this  relation predicts that $P_c$ increases as well. This implies that waves going in the \textit{opposite} direction should be unstable. Under these irreconcilable conditions, the cosmic ray pressure gradient flattens, there are no waves, and cosmic rays do not exchange energy or parallel momentum with the ambient medium.

Bottlenecks typically take at least one Alfven crossing time to form, so their steady state structure may not always be achieved, but they imply that self confinement, and the heating and momentum
transfer that accompany it, are quite intermittent.  Since much of the diffuse gas in the Universe is clumpy and astrophysical magnetic fields are usually tangled, this intermittency might be the generic state. This may be a confounding issue, for example, in treatments
of cosmic ray propagation and cosmic ray heating in clusters of galaxies, where the magnetic geometry is usually assumed to be simple \citep{LoewensteinZweibelBegelman1991,guooh08,wieneretal13b,jacobpfrommer17a,jacobpfrommer17b,wienerzweibel19}.

Because cosmic ray heating and momentum transfer are important in so many astrophysical systems, we must ask whether drift anisotropy is the only path to self confinement. 
 In this paper we develop a complementary theory for cosmic ray - thermal gas coupling mediated by hydromagnetic waves. The theory is based on an instability similar to the drift instability, but the energy source is pressure anisotropy. 
 
Pressure anisotropy instability was studied previously by \cite{lazarianberesnyak06}, hereafter LB06, and by \cite{yanlazarian11}, who investigated it as a mechanism for cosmic ray
self confinement and an energy sink for interstellar turbulence, and was recently studied numerically by \cite{lebiga2018}. 
turbulence and acted as a sink for turbulent energy, 
We instead consider a case in which the instability is driven by an input of cosmic ray energy itself.  This is the most direct conceptual analog of heating due to streaming down the cosmic ray pressure gradient.

The main outcome of our work is that for global or galactic scale processes,  pressure anisotropy instability is much weaker than drift anisotropy as a mechanism for cosmic ray self confinement and plasma heating. We find that under otherwise similar conditions, the spatial diffusivity resulting for pressure anisotropy is a factor of order $c/v_A$ larger than the
diffusivity resulting for drift anisotropy, and the heating rate associated with pressure anisotropy is lower than that due to drift anisotropy by a factor of order $v_A/c$. The underlying reason is that when scattering is due to drift anisotropy instability, it works against the tendency for cosmic rays to stream at the speed of light while the characteristic speed associated with driving pressure anisotropy is the magnetoacoustic speed in the thermal gas. Therefore, a much lower level of turbulent fluctuations is required to maintain marginal stability
against pressure anisotropy. The lower energy density in fluctuations translates to a lower level of heating. Phenomena on smaller spatial scales can drive anisotropy harder, leading to stronger scattering and possibly a significant level of heating.

  In \S\ref{s:formulation} we pose the simplest possible problem that brings out the relevant effects: expansion of a magnetic flux tube due to spatially uniform injection of cosmic ray pressure. In \S\ref{s:instability} we discuss the pressure anisotropy instability. In
\S\ref{s:heating} we derive an expression for the heating rate that results from instability due to cosmic ray driven flux tube expansion, and in \S\ref{s:Implications} we discuss the implications of the instability for cosmic ray self confinement and estimate the spatial diffusion coefficient. In \S \ref{s:Application} we apply the theory to the problem of a bottleneck formed between a galactic halo cloud and a
galactic disk \citep{wieneretal2019}.
Section \S\ref{s:Summary} is a summary, speculation on how drift and pressure anisotropy instability might combine, and conclusion.

\section{Formulation of the Problem}\label{s:formulation}
\subsection{Macroscopic Dynamics}
Here we develop a simple model in which  input of cosmic ray energy drives expansion of the ambient gas, which weakens the magnetic field and causes cosmic ray pressure anisotropy. The anisotropy   triggers instabilities that scatter the cosmic rays and heat the gas, leading to a relationship between the energy injection rate, the rate of doing work, and the rate of heating. An example of a situation in which these events might occur is shown in Figure 1a, while the simplified problem we solve is shown in Figure 1b.
\begin{figure}
\begin{center}
\includegraphics[height=40mm]{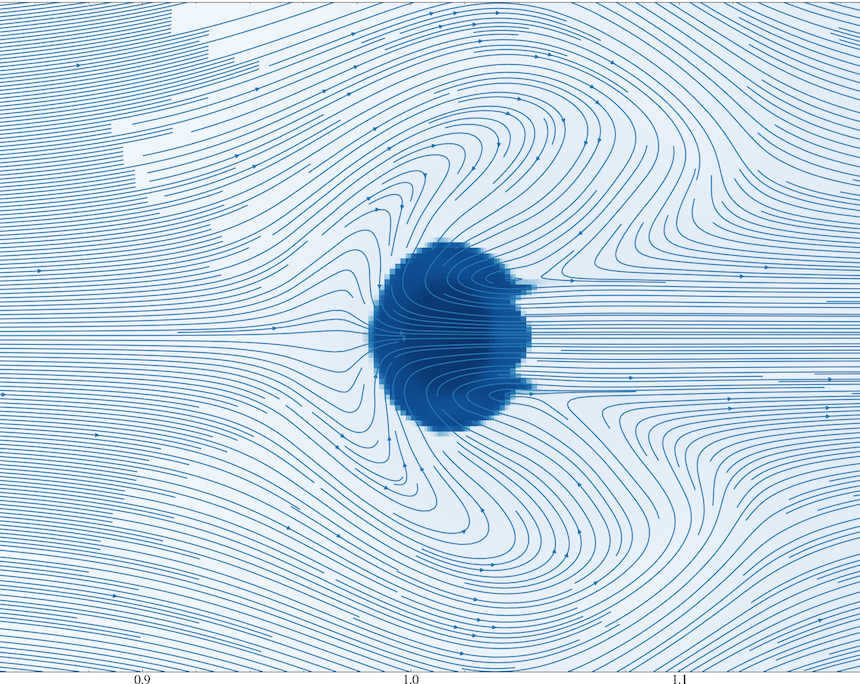}
\includegraphics[height=40mm]{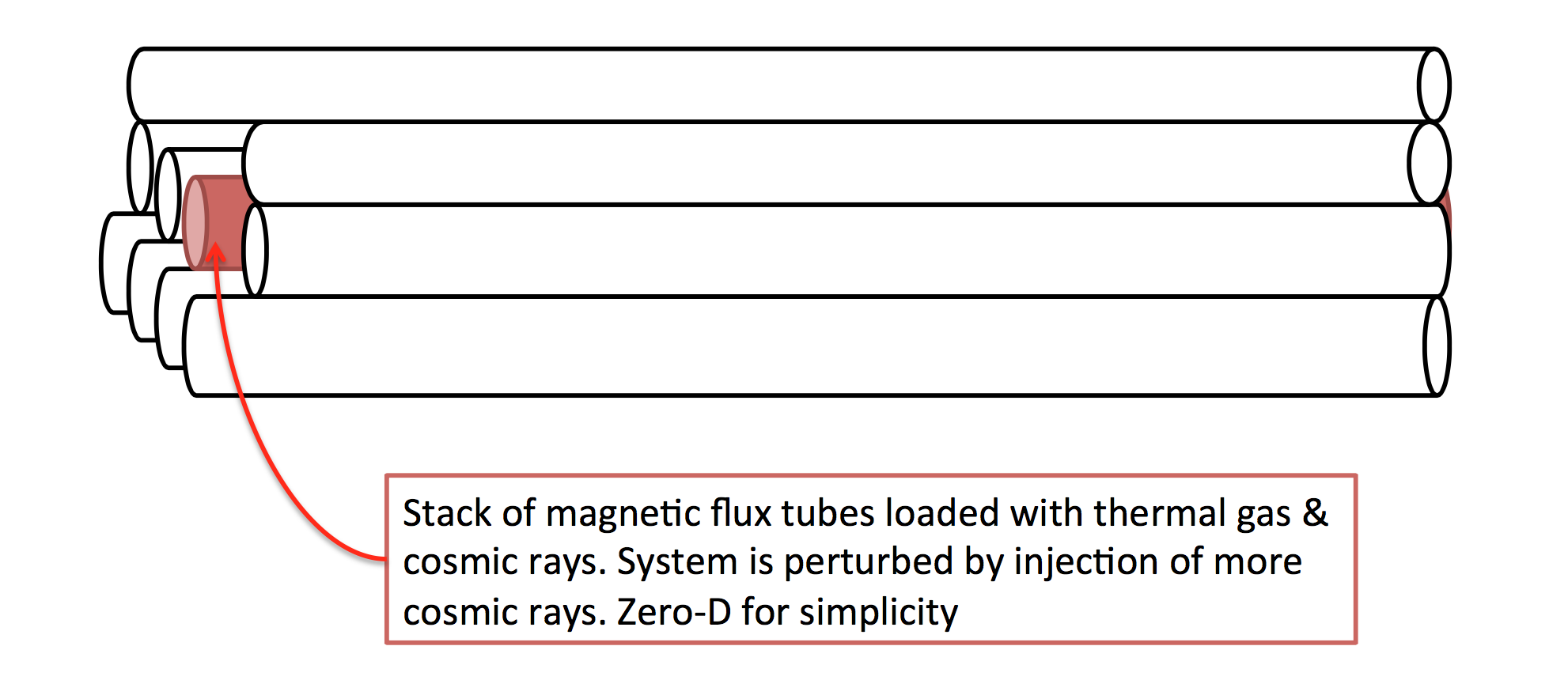}
\end{center}
\caption{Top: Distortion of a weak magnetic field when cosmic rays propagating away from a source (left boundary) encounter a denser  cloud inside which the Alfv\'en speed drops. In the absence of scattering, the cosmic ray pressure will become highly anisotropic. From simulations that led to, but are not included in, \cite{wieneretal2019}. Bottom: Sketch of the
model on which the calculations in this paper are based. Cosmic rays are injected uniformly onto a thin magnetic flux tube, which expands to maintain pressure equilibrium with its
surroundings.}
\end{figure}

Consider a uniform medium with a background magnetic field $\mbfB$, thermal gas pressure $P_g$, and cosmic ray pressure $P_c$. This equilibrium system is perturbed by injecting cosmic rays  onto a magnetic flux tube, or bundle of field lines, of radius $R$, at the same rate $\Delta\dot P_c$, everywhere along the flux tube. In a more realistic situation the cosmic
ray source is likely to be localized to a small region of the flux tube and the cosmic rays  stream away along the field lines as well as causing the magnetic field to locally
expand. This corresponds to the standard ``streaming" picture;  here we bring out the transverse dynamics\footnote{If the injection profile of the cosmic rays is sharp, producing
strong cross-field gradients, this itself can be unstable \citep{riquelmespitkovsky2010}, but we ignore that possibility here}.

The flux tube responds to cosmic ray injection by expanding perpendicular to its major axis. We approximate the speed of expansion  by the magnetoacoustic speed $C_{ma}$
\begin{equation}\label{eq:cma}
C_{ma}\equiv\left(\frac{\gamma_gP_g+\gamma_cP_c+\gamma_mP_m}{\rho_g}\right)^{1/2},
\end{equation}
where $\gamma_g$ and $\gamma_p$ are the thermal and cosmic ray polytropic indices, $P_m\equiv B^2/(8\pi)$ is the magnetic pressure and $\gamma_m=2$ for transverse expansion. In writing eqn. (\ref{eq:cma}) we have made several simplifying assumptions. We have replaced the perpendicular cosmic ray pressure $P_{c\perp}$ by the isotropic pressure $P_c$; assuming that scattering by waves keeps the pressure anisotropy near marginal stability, this
is accurate to order $v_A/c$. 
We have also  assumed that the effects of heating and cooling on the thermal gas can be subsumed into an effective polytropic index $\gamma_g$; a simplification which is common in interstellar gas dynamics. 

The characteristic  expansion timescale for the tube is $\tau_D\sim R/C_{ma}$ while the  total pressure changes on a timescale $\tau_P\sim (P_g+P_c+P_m)/\Delta\dot P_c$. If $\tau_D/\tau_P\ll 1$, the tube maintains pressure equilibrium with its surroundings; this sets an upper limit on the radius $R$. By the same assumptions made in deriving eqn.
(\ref{eq:cma}),
\begin{equation}\label{eq:dotP}
C_{ma}^2\frac{d\rho_g}{dt}=-\Delta\dot P_{c},
\end{equation}
or, because $B/\rho_g$ is constant for uniform, transverse expansion
\begin{equation}\label{eq:dotn}
\frac{\dot\rho_g}{\rho_g}=\frac{\dot B}{B}=-\frac{\Delta\dot P_c}{\rho_gC_{ma}^2}\equiv -\frac{1}{\tau_B},
\end{equation}
where we have introduced the magnetic field timescale $\tau_B$ for later use.  
From the First Law of Thermodynamics, the rate $\Delta\dot W_c$ at which the cosmic rays do work is
\begin{equation}\label{eq:dotWc}
\Delta\dot W_c=\frac{P_c\Delta\dot P_c}{\rho_gC_{ma}^2}=-P_c\frac{\dot B}{B}=\frac{P_c}{\tau_B}.
\end{equation}
In\S\ref{s:heating} we derive a rate of heating in terms of the rate of doing work.

\subsection{Microscale Response}
As the flux tube expands, the cosmic ray momentum distribution becomes anisotropic with respect to $\mbfB$. If the magnetic field changes slowly relative to the cosmic ray gyration frequency and there is no scattering, the particle motion is adiabatic and two orbital properties, the magnetic moment $p^2(1-\mu^2)/B$ and longitudinal action $p\mu$ can be treated as constant. Here, $p$ is the magnitude of the momentum and $\mu\equiv\mbfp\cdot\mbfB/pB$ is the
cosine of the pitch angle. The evolution of the cosmic ray phase space distribution function $f(\mbfp, t)$ is given by
\begin{equation}\label{eq:firstkinetic}
\frac{\partial f}{\partial t} +\frac{dp}{dt}\frac{\partial f}{\partial p} + \frac{d\mu}{dt}\frac{\partial f}{\partial\mu}=\Delta\dot f(p,t),
\end{equation}
or, assuming adiabatic motion,
\begin{equation}\label{eq:fpa}
\frac{\partial f}{\partial t}+\frac{\dot B}{B}\frac{\left(1-\mu^2\right)}{2}\left(p\frac{\partial f}{\partial p}-\mu\frac{\partial f}{\partial\mu}\right)=\Delta\dot f(p,t), 
\end{equation}
where $\Delta\dot f(p,t)$ is the phase space counterpart of $\Delta\dot P_c$ and we assume particles are injected isotropically (see \cite{lichko17} for a more general version
of eqn. \ref{eq:fpa} that includes shearing as well as compression).

The evolution of $f$  can be followed more easily if we expand it as a Legendre series in $\mu$ 
\begin{equation}\label{eq:series}
f(p,\mu, t)=\Sigma_{l=0}^{\infty}f_l(p,t)P_l(\mu),
\end{equation}
where
\begin{equation}\label{eq:fl}
f_l(p,t)\equiv\frac{2l+1}{2}\int_{-1}^{1}f(p,\mu,t)P_l(\mu)d\mu.
\end{equation}
It can be shown from well known relations between Legendre functions that the solution of eqn.(\ref{eq:fpa}) contains all harmonics of even order $l=2n$,  even if $f$ is initially isotropic.
However, we show in \S\ref{s:instability} that the magnitude  of the anisotropy is capped at a value of order $v_A/c$. Therefore, the primary driving term is isotropic, and we approximate eqn. (\ref{eq:fpa}) by
\begin{equation}\label{eq:fpaapr}
\frac{\partial f}{\partial t}+\frac{\dot B}{3B}\left(P_0(\mu)-P_2(\mu)\right)p\frac{\partial f_0(p,t)}{\partial p}=\Delta\dot f(p,t).
\end{equation}
According to eqn. (\ref{eq:fpaapr}), if $f$ is initially isotropic, it only generates a $P_2$ anisotropy. In our problem $\dot B/B < 0$, so assuming $df_0/dp < 0$, the anisotropy is
positive.

The physical significance of $f_2$ becomes apparent if we compute the  pressure anisotropy 
\begin{equation}\label{eq:f2}
\begin{split}
\Delta P_c\equiv P_{\parallel}-P_{\perp}\equiv\int fpv\left(\mu^2-\frac{1-\mu^2}{2}\right)p^2dpd\mu=\\ \int fpvP_2(\mu)p^2dpd\mu=
\frac{2}{5}\int f_2pvp^2dp.
\end{split}
\end{equation}
Equation (\ref{eq:f2}) shows that $f_2$ is a direct measure of pressure anisotropy. Strictly speaking, it is $P_{\perp}$ that drives the expansion of the flux tube, but for the small anisotropy expected here,  this is a correction which we can ignore.

\section{Pressure Anisotropy Instability}\label{s:instability}

We are interested in instabilities of hydromagnetic waves which are driven by gyroresonant particles. A particle gyroresonates with a circularly polarized wave of frequency $\omega$ and wavenumber $k_{\parallel}$ parallel to $\mbfB$ if its parallel velocity $v_{\parallel}$ and relativistic gyrofrequency $\Omega = \Omega_0/\gamma$ (where $\Omega_0$ is the nonrelativistic gyrofrequency
and $\gamma$ is the Lorentz factor) satisfy the condition that the wave frequency Doppler shifted to  the particle frame matches the
cyclotron frequency 
\begin{equation}\label{eq:gyro}
\omega-k_{\parallel}v\mu\pm\Omega=0.
\end{equation}
Here the $\pm$ signs refer to right and left circular polarization respectively. We assume throughout the paper  that the cosmic rays and interstellar ions are both protons, but the analysis can be generalized to other ion species. 

Since  $\vert\omega/k_{\parallel}v\vert\sim v_A/c$, we can drop $\omega$ in evaluating the resonance condition, unless $\vert\mu\vert\ll 1$. The resonant $\mu$, $\mu_r$, is then
\begin{equation}\label{eq:mur}
\mu_r=\pm\frac{\Omega}{k_{\parallel}v}=\pm\frac{m_i\Omega_0}{k_{\parallel}p}\equiv\pm\frac{p_1}{p},
\end{equation}
where $p_1\equiv m_i\Omega_0/k_{\parallel}$ is the minimum momentum which can resonate with a  wave of parallel wavenumber $k_{\parallel}$.

 \cite{kulsrudpearce69} derived the growth rate for hydromagnetic waves destabilized by cosmic rays, and showed  that waves propagating parallel to $\mbfB$  grow faster than oblique waves. We  use the Kulsud and Pearce expression and assume parallel propagation, but consider the two
 senses of circular polarization separately rather than linear polarization as assumed by them. The growth rate of parallel propagating, circularly polarized Alfven waves, is 
\begin{equation}\label{eq:Gamman}
\Gamma_c=\frac{\pi^2q^2}{2}\frac{v_A^2}{c^2}\int p^2dpd\mu v(1-\mu^2)\delta(\omega-kv\mu\pm\Omega)A(f,\omega,k),
\end{equation}
 where we have suppressed the subscript on $k$ because we now take $k=k_{\parallel}$. The anisotropy functional $A$  is

 \begin{equation}\label{eq:A}
A\left[f,\omega,k\right]\equiv\frac{\partial f}{\partial p}+\left(\frac{kv}{\omega}-\mu\right)\frac{1}{p}\frac{\partial f}{\partial\mu}.
 \end{equation}
 %
From now on  we will  drop the $\mu$ term multiplying $\partial f/\partial\mu$ relative to $kv/\omega\sim c/v_A$ in $A$. 

Integrating eqn. (\ref{eq:Gamman}) over $\mu$,
using eqn. (\ref{eq:mur}), and rearranging the prefactor gives
\begin{equation}\label{eq:Gammaint}
\Gamma_c=\frac{\pi}{8}\frac{\Omega_0}{n_i}\int_{p_1}p_1dp\left(p^2-p_1^2\right)A[f,\omega,k].
\end{equation} 
 Instability requires $A>0$ in at least some
 part of phase space. Since $\partial f_0/\partial p < 0$ for typical cosmic ray distributions, the source of instability must be in the anisotropy term. However, due to the factor
 of $kv/\omega\sim c/v_A$ multiplying $\partial f/\partial\mu$, the anisotropic part of $f$ need only be of order $v_A/c$ relative to the isotropic part for the wave to be unstable.

 The  classical streaming instability occurs for drift anisotropy of the form
\begin{equation}\label{eq:driftanisotropy}
f(p,\mu)=f_0(p)-\alpha p\frac{df}{dp}P_1(\mu).
\end{equation}
It can be shown from the Lorentz invariance of $f$ that the distribution function given eqn. (\ref{eq:driftanisotropy})  is isotropic in a frame moving with speed $\alpha c$, up to factors of order $\alpha^2$.  Assuming the particles are ultrarelativistic ($v\sim c$), the parameter $\alpha$ is directly related to the bulk drift, or streaming velocity by
\begin{equation}\label{eq:vD}
v_D\equiv\frac{1}{n_c}\int v\mu f(p,\mu)p^2dpd\mu = \alpha c.
\end{equation}
Substituting eqn. (\ref{eq:driftanisotropy}) into eqn. (\ref{eq:A}) and using eqn. (\ref{eq:vD}) gives
\begin{equation}\label{eq:Ad}
A[f,\omega,k]\equiv A_d[f,\omega,k]=\frac{df_0}{dp}\left(1-\frac{v_D}{v_A}\right),
\end{equation}
where we have taken $\omega/k = v_A$. Then, substituting eqn. (\ref{eq:driftanisotropy}) into eqn. (\ref{eq:Gammaint}) gives $\Gamma_{cd}$, the growth rate of the streaming instability 
\begin{equation}\label{eq:Gammacd}
\Gamma_{cd}=\frac{\pi}{8}\Omega_0{\mathcal{C}}\frac{n_c(>p_1)}{n_i}\left(\frac{v_D}{v_A}-1\right)
\end{equation}
where ${\mathcal{C}}$ is a spectrum - dependent constant of order unity and $n_c(>p_1)$ is the number density of cosmic rays with momentum $p > p_1$, which are the only cosmic
rays which can resonate with the wave.

In the case of pure pressure anisotropy considered in \S\ref{s:formulation}, there are no gradients along the magnetic field, so the series (\ref{eq:series}) contains only terms with even $l$. Therefore,
$\partial f/\partial\mu$ is an odd function of $\mu$. If $\partial f/\partial\mu > 0$ $(<0)$, only waves propagating in the positive (negative) direction can be unstable. But if particles
of some particular $\mu_r$ resonate with a wave of one polarization, particles with $\mu=-\mu_r$ resonate with a wave of the opposite polarization. Therefore, if a circularly polarized wave  is unstable, the wave with the opposite circular polarization propagating in the opposite direction is also unstable, and has the same growth rate. Therefore, cosmic rays confined by pressure anisotropy
instability would be advected by the thermal gas, rather than streaming relative to it at the Alfv\'en speed.

We make the \textit{ansatz}
\begin{equation}\label{eq:pressureanisotropyform}
f(p,\mu)=f_0(p)-\zeta p\frac{df_0}{dp}P_2(\mu)
\end{equation}
(eqn. \ref{eq:fppas} provides some support for this form of $f_2$). The parameter $\zeta$ is related to the pressure anisotropy by
\begin{equation}\label{eq:zeta}
\zeta = \frac{5}{12}\frac{\Delta P_{c}}{P_c},
\end{equation}
where $P_c\equiv (2P_{c\perp}+P_{c\parallel})/3$ is the scalar pressure.  Substituting eqn. (\ref{eq:pressureanisotropyform}) into eqn. (\ref{eq:A}) and approximating $\partial f/\partial p$ by $\partial 
f_0/\partial p$ gives
\begin{equation}\label{eq:Af2}
A_{pa}\left[f,\omega,k\right]=\frac{\partial f_0}{\partial p}\left(1-\frac{3\zeta kv\mu}{\omega}\right)
\end{equation}
which is to be evaluated at $\mu_r$. From eqn (\ref{eq:mur}) and assuming $\zeta > 0$, as it is for our problem, we see that if the polarization is chosen such that the
resonant particles are traveling in the same direction as the wave, the wave can be unstable, while the opposite sense of circular polarization is always damped. 
That is,  right circularly polarized waves propagating in the positive direction  and left circular polarized waves propagating in the negative direction are the
only possible unstable ones (Similar arguments show that in a compressing flux tube with $\zeta < 0$,  the instabilities are positive propagation direction/left circular polarization and negative propagation direction/right circular polarization). On the other hand, linearly polarized waves are neutrally stable, as the
growth and damping contributions from the two circularly polarized components cancel one another. 

In the important case of a power law cosmic ray distribution $f_0(p)\propto p^{-a}$,
eqns. (\ref{eq:Gammaint}) and (\ref{eq:zeta}) yield for the growth rate for pressure anisotropy instability

\begin{equation}\label{eq:Gammanpower}
\Gamma_{cpa}=\frac{\pi}{8}{\mathcal{C}}\Omega_0\frac{n_c(>p_1)}{n_i}\left[\frac{5 a(a-2)}{4(a^2-1)}\frac{c}{v_A}\frac{\Delta P_c}{P_c}-1\right],
\end{equation}
%
%
with ${\mathcal{C}}=(a-3)/(a-2)$.
In the strong anisotropy limit, which corresponds to dropping the ``1" in the square bracket on the right
hand side of eqn. (\ref{eq:Gammanpower}), our expression for the instability growth rate agrees with the strong anisotropy limit given in  LB06. For the local interstellar cosmic ray
spectrum with $a\sim 4.7$, ${\mathcal{C}}=0.63$, and the threshold anisotropy for instability is  $\zeta > 0.55$, or $\Delta P_c/P_c\sim 1.3v_A/c$.

\section{Calculation of the Heating Rate}\label{s:heating}

Our tool for calculating the rate at which cosmic rays heat the thermal plasma will be the Fokker-Planck equation, which is the Vlasov equation plus scattering terms. As is usual in quasilinear diffusion theory, we assume the waves are small amplitude and have random phases. We consider only resonant  wave - particle interactions  with parallel propagating Alfv\'en waves. We begin with a brief review of heating when the waves are generated by drift anisotropy and then calculate heating when the waves are generated by pressure anisotropy, which is the main contribution of
this paper.

\subsection{Heating due to drift anisotropy}\label{ss:Drift}

We assume $f$ has a spatial gradient along the background magnetic field $\mbfB=\hat s B$ such that the cosmic rays drift  toward positive $s$. The Fokker-Planck equation is
\begin{equation}\label{eq:fpd}
\frac{\partial f}{\partial t}+v\mu\frac{\partial f}{\partial s}=\frac{\partial F_{\mu}}{\partial\mu} + \frac{1}{p^2}\frac{\partial}{\partial p}\left(p^2F_p\right),
\end{equation}
where $F_{\mu}$ and $F_p$ are the components of the diffusive flux, which can be written in terms of components of the momentum
space diffusion tensor $\mbfD$ as
\begin{equation}\label{eq:fmu}
F_{\mu}=D_{\mu\mu}\frac{\partial f}{\partial\mu}+D_{\mu p}\frac{\partial f}{\partial p},
\end{equation}
\begin{equation}\label{eq:fp}
F_p=D_{p\mu}\frac{\partial f}{\partial\mu}+D_{pp}\frac{\partial f}{\partial p}.
\end{equation}
In the case of drift anisotropy, only waves propagating toward positive $s$ are present, but with both signs of circular polarization (which we assume have equal
intensity). The components of  $\mbfD$ are \citep{Schlickeiser89}
\begin{equation}\label{eq:Dmm}
D_{\mu\mu}=\frac{\nu\left(1-\mu^2\right)}{2},
\end{equation}
\begin{equation}\label{eq:Dmup}
D_{\mu p}=D_{p\mu}=\frac{pv_A}{v}D_{\mu\mu},
\end{equation}
\begin{equation}\label{eq:Dpp}
D_{pp}=\left(\frac{pv_A}{v}\right)^2D_{\mu\mu},
\end{equation}
where $\nu$ is the pitch angle scattering frequency, which is related to the spectral magnetic energy density of resonant waves $W_k$ by
\begin{equation}\label{eq:nu}
\nu(p,\mu)=\Omega\frac{8\pi kW_k}{B^2}.
\end{equation}
Equations (\ref{eq:Dmm} - \ref{eq:Dpp}) are valid for small amplitude, parallel propagating Alfv\'en waves of arbitrary polarization
traveling parallel or antiparallel to $\mbfB$.

Substituting eqns. (\ref{eq:Dmm}) and (\ref{eq:Dmup}) into eqn. (\ref{eq:fmu}) gives
\begin{equation}\label{eq:fmuA}
F_{\mu}=\frac{\nu\left(1-\mu^2\right)}{2}\frac{pv_A}{v}A(f,kv_A,k),
\end{equation}
where $A$, the functional introduced in eqn. (\ref{eq:A}), is fundamental to the criterion for instabiity (eqn. \ref{eq:Gamman}). Likewise,
\begin{equation}\label{eq:fpA}
F_p=\frac{\nu\left(1-\mu^2\right)}{2}\left(\frac{pv_A}{v}\right)^2A(f,kv_A,k).
\end{equation}
The implications of the close relationship between momentum space diffusion and wave excitation (or damping) will shortly become clear.

In some early studies of  self confinement by drift anisotropy, such as \cite{kulsrudpearce69} and \cite{Skilling71,Skilling75}, the Fokker-Planck equation is solved in a frame moving with the waves, in which case the cosmic ray scattering
is purely elastic, with $D_{\mu\mu}$ the only nonzero component of the diffusion tensor. Because we are interested in energy exchange between the cosmic rays and the background, and because pressure anisotropy excites waves propagating in both directions, we work in the rest frame of the plasma.

We  solve eqn. (\ref{eq:fpd}) under the assumption that the scattering mean free path $\lambda\equiv v/\nu$ is short compared to the gradient lengthscale $L_c\equiv\vert f_c/(\partial f_c /\partial s)\vert$. and the cosmic ray streaming anisotropy is ${\mathcal{O}}(v_A/c)\ll 1$. We then keep only the second term on the LHS of eqn. (\ref{eq:fpd}), and only the first term on the RHS. Integrating once with respect to $\mu$ and using eqns. (\ref{eq:Dmm}) and (\ref{eq:Dmup}) gives a relationship between the spatial gradient of $f$, the collision frequency $\nu$, and $pv_AA/v$
%
\begin{equation}\label{eq:fpds}
\frac{pv_A}{v}\left(\frac{\partial f}{\partial p}+\frac{v}{v_Ap}\frac{\partial f}{\partial\mu}\right)=-\frac{v}{\nu}\frac{\partial f}{\partial s}.
\end{equation}
If we replace $\partial/\partial s$ by a gradient length scale $L^{-1}$, we see that the ratio of the anisotropic to the isotropic part of $f$ is of order $c/(\nu L)$.

Next, we derive an energy equation from eqn. (\ref{eq:fpd}) by multiplying by particle energy $\epsilon$ and integrating over phase space. The result is
\begin{equation}\label{eq:ed}
\frac{\partial U_c}{\partial t}+\frac{\partial F_{c\eps}}{\partial s}=-\int vF_pp^2dpd\mu,
\end{equation}
where
\begin{equation}\label{eq:Uc}
U_c\equiv\int\eps f p^2dpd\mu,
\end{equation}
\begin{equation}\label{eq:Fe}
F_{c\eps}\equiv\int v\mu\eps fp^2dpd\mu
\end{equation}
are the cosmic ray energy density and energy flux vector, respectively. The right hand side of eqn. (\ref{eq:ed}) represents energy transfer between cosmic rays and waves due to scattering. It is shown in the Appendix that this can be written in terms on the wave energy densities and growth rates as
\begin{equation}\label{eq:vfp}
H_d\equiv \int vF_pp^2dpd\mu = 2\int\Gamma_c(k)W_kdk,
\end{equation}
which was first shown for gyroresonant instability by \cite{kennel66}. Here, we evaluate the energy transfer term directly using eqn. (\ref{eq:fpds})
\begin{equation}\label{eq:etd}
H_d=-\int v^2\frac{pv_A}{v}\frac{\left(1-\mu^2\right)}{2}\frac{\partial f_c}{\partial s}p^2dpd\mu=-v_A\frac{\partial P_c}{\partial s},
\end{equation}
where for streaming toward increasing $s$, $\partial P_c/\partial s < 0$.  

Equation (\ref{eq:etd}) agrees with the standard expression for the collisionless heating rate, and is notable for its
simplicity and lack of any explicit dependence on wave properties. It does, however, have an implicit dependence in that the cosmic ray pressure gradient
is determined by transport, and the transport is determined, in part, by the degree of scattering. But as eqns. (\ref{eq:fpds})  and (\ref{eq:Gamman}) show, the scattering rate is linked to the wave growth rate. In a steady state, wave growth  is balanced by wave damping by the thermal background. So ultimately, the pressure
gradient and heating rate are determined by the properties of the cosmic ray source and the background gas through which the cosmic rays stream. In particular, if the waves are heavily damped, the cosmic ray anisotropy must be large to overcome wave damping. This corresponds to a long scattering mean free path or a large diffusion coefficient, which flattens the cosmic ray pressure gradient and reduces the rate at which it can heat and do work on the thermal gas.

\subsection{Heating due to pressure anisotropy}\label{ss:Pressure}

Now, we attempt to derive an expression for the heating rate due to pressure anisotropy instability that is as compact as eqn. (\ref{eq:etd}). Under the terms of our problem, eqn. (\ref{eq:fpd}) is replaced by eqn. (\ref{eq:fpaapr}) plus momentum space diffusion terms
\begin{equation}\label{eq:fppa}
\begin{split}
\frac{\partial f}{\partial t}+\frac{\dot B}{3B}\left(P_0(\mu)-P_2(\mu)\right)p\frac{\partial f_0}{\partial p}=\Delta\dot f(p) \\
+\frac{\partial F_{\mu}}{\partial\mu} + \frac{1}{p^2}\frac{\partial}{\partial p}\left(p^2F_p\right),
\end{split}
\end{equation}

As discussed below eqn. (\ref{eq:Af2}),  waves of opposite circular polarization propagate in opposite directions, and particles resonate with waves traveling in the same direction they are (we assume the
waves propagate in both directions with equal intensity). Accordingly, eqns. (\ref{eq:Dmm}) and (\ref{eq:Dpp}) respectively can be used for $D_{\mu\mu}$ and $D_{pp}$, but
\begin{equation}\label{eq:Dmppap}
D_{\mu p}=D_{p \mu}=\frac{pv_A}{v}D_{\mu \mu};\mu > 0,
\end{equation}
\begin{equation}\label{eq:Dmppam}
D_{\mu p}=D_{p \mu}= - \frac{pv_A}{v}D_{\mu \mu};\mu < 0.
\end{equation}
The discontinuity in the off diagonal terms $D_{\mu p}$ and $D_{p \mu}$ is only apparent, as $\nu\equiv 0$ for $\mu=0$ for any distribution of waves with a short wavelength cutoff. Scattering mechanisms which supplement pitch angle scattering at small $\mu$, such as mirroring, have been proposed \citep{FeliceKulsrud01} but we ignore them here; we have also dropped terms of order $v_A/c$ in the diffusion tensor which remove the singularity at $\mu = 0$ \citep{Schlickeiser89}. Importantly,  because for pressure anisotropy $\partial f_c/\partial\mu$ is odd in $\mu$, $F_{\mu}$ is odd in $\mu$ while $F_p$ is even.

We solve for the anisotropy driven by the time varying magnetic field by multiplying eqn. (\ref{eq:fppa}) by $P_2(\mu)$ and integrating over $\mu$, making the same assumptions about the ordering of terms we made in deriving eqn. (\ref{eq:fpds}) from eqn. (\ref{eq:fpd}); we drop $\partial f/\partial t$, replace $f$ by $f_{0}$ in the $\dot B/B$ term, and
drop $F_p$ but keep $F_{\mu}$ on the right hand side. The result is
\begin{equation}\label{eq:fppas}
\begin{split}
\frac{pv_A}{v}\int_0^1\nu\mu\left(1-\mu^2\right)\left(\frac{\partial f}{\partial p}+\frac{v}{v_Ap}\frac{\partial f}{\partial\mu}\right)d\mu= \\ \frac{pv_A}{v}  \int_{0}^{1}\nu\mu\left(1-\mu^2\right)A[f,\omega,k]d\mu \\=\left(\frac{2}{45}\right)\frac{\dot B}{B}p\frac{\partial f_0}{\partial p}.
\end{split}
\end{equation}

The left hand side of eqn. (\ref{eq:fppas}) is proportional to the anisotropy factor $A$ while the right hand side of eqn. (\ref{eq:fppas}) is positive, as expected for anisotropy driven instability.  However, whereas eqn. (\ref{eq:fpds}) gives $A$ directly as a function of $\nu$ and $p$, eqn. (\ref{eq:fppas}) involves an integral of $A$ with $\nu$. The more important
difference between eqns. (\ref{eq:fpds}) and (\ref{eq:fppas}) is that the timescale on the right hand side of eqn. (\ref{eq:fpds}) is the light travel time $L_c/c$, while the timescale on
the right hand side if eqn. (\ref{eq:fppas}) is the flux tube expansion timescale $\vert B/\dot B\vert$. 
We discuss the implications of eqn. (\ref{eq:fppas}) for cosmic ray self confinement in \S\ref{s:Implications}.

We derive an energy equation analogous to eqn. (\ref{eq:ed}) by multiplying eqn. (\ref{eq:fppa}) by $\eps$ and integrating over momentum space. The result is
\begin{equation}\label{eq:epa}
\frac{\partial U_c}{\partial t}-\frac{\dot B}{B}\left(U_c+P_{c\perp}\right)=\Delta\dot U_c-\int vF_pp^2dpd\mu.
\end{equation}

The first term on the right hand side of eqn. (\ref{eq:epa}) is the cosmic ray energy injection rate corresponding to the cosmic ray pressure source.
If the pressure were isotropic and we substituted $\dot\rho_g/\rho_g$ for $\dot B/B$, the left hand side of eqn. (\ref{eq:epa}) would be in standard form for describing adiabatic expansion.  Since in the absence of collisions, $P_{\perp}$ decreases slightly faster than $P_c$ itself, anisotropy slows the rate of energy loss. If we use the identity
\begin{equation}\label{eq:identity}
P_{c\perp}=P_c+\frac{P_{c\perp}-P_{c\parallel}}{3}
\end{equation}
then eqn. (\ref{eq:epa}) can be written as
\begin{equation}\label{eq:epa2}
\begin{split}
&\frac{\partial U_c}{\partial t}-\frac{\dot B}{B}\left(U_c+P_{c}\right)=\\
&\frac{\dot B}{3B}\left(P_{c\perp}-P_{c\parallel}  \right) +\Delta\dot U_c-\int vF_pp^2dpd\mu.
\end{split}
\end{equation}
Since pressure anisotropy is reversed in an increasing magnetic field, the anisotropy term on the right hand side of  eqn. (\ref{eq:epa2}) is positive whatever the sign of $\dot B/B$.
It shows that  anisotropy reduces the rate of energy loss in a decreasing magnetic field and increases the rate of energy gain in an increasing field. In both cases, this is because
the parallel momentum is fixed. Although the anisotropy term formally resembles gyroviscous heating \citep{Kunzetal2011}, it is not a true heating process because it is completely reversible.

Energy transfer to waves, however, is a true energy loss process. To evaluate it,  we write out the diffusion term explicitly, using the even parity of $F_p$ noted below eqn. (\ref{eq:fppas}) and eqn. (\ref{eq:A}). This gives
\begin{equation}\label{eq:etda}
\begin{split}
&\int vF_pp^2dpd\mu= \\
&\int p^2dpv\left(\frac{pv_A}{v}\right)^2\int_0^1\nu\left(1-\mu^2\right)A(f,kv_A,k)d\mu.
\end{split}
\end{equation}
We know from eqn. (\ref{eq:fppas}) that
\begin{equation}\label{eq:ineq}
\begin{split}
\left\vert\int_0^1\nu\left(1-\mu^2\right)\left(\frac{\partial f_c}{\partial p}+\frac{v}{v_Ap}\frac{\partial f_c}{\partial\mu}\right)d\mu\right\vert\ge \\
\left\vert\int_0^1\nu\mu\left(1-\mu^2\right)\left(\frac{\partial f_c}{\partial p}+\frac{v}{v_Ap}\frac{\partial f_c}{\partial\mu}\right)d\mu\right\vert \\
=\left\vert\frac{2}{45}\frac{v}{v_Ap}\frac{\dot B}{B}p\frac{\partial f_c}{\partial p}\right\vert.
\end{split}
\end{equation}
(where the first inequality simply follows from $\mu\le 1$). Therefore, we have a lower bound on the magnitude of cosmic ray heating
\begin{equation}\label{eq:hbound}
\begin{split}
\left\vert\int vF_pp^2dpd\mu\right\vert\equiv H_{pa}\ge\left\vert\frac{2}{45}v_A\frac{\dot B}{B}\int p^4\frac{df_c}{dp}\right\vert \\
=\left\vert\frac{4}{15}\frac{v_A}{c}\frac{\dot B}{B}P_c\right\vert=\frac{4}{15}\frac{v_A}{c}\frac{P_c\Delta\dot P_c}{\gamma_cP_c+\gamma_mP_m+\gamma_gP_g},.
\end{split}
\end{equation}
where in the last step we have used eqn. (\ref{eq:dotn}) and (\ref{eq:cma}). Equation (\ref{eq:hbound}) is only a lower bound because our analysis does not give the functional form of $\nu$. Experimentation with various trial functions for $\nu$
suggests that eqn. (\ref{eq:hbound}) is unlikely to underestimate the heating by more than a factor of 2. 

We can generalize eqn. (\ref{eq:hbound}) by comparing the rates at which the cosmic rays are heating their environment to the rate at which they are doing work on
it, which from eqn. (\ref{eq:dotWc}) is $-P_c\dot B/B$. Therefore, the rate of heating is the rate of work multiplied by a factor of order $v_A/c$.

\subsection{Comparison of  Drift and Pressure Anisotropy Heating}\label{ss:Comparison}

The rate of cosmic ray heating due to streaming anisotropy (eqn. \ref{eq:etd}) is proportional to $\nabla_{\parallel}P_c$, while the heating rate due to pressure anisotropy (\ref{eq:hbound}) is proportional to $\dot P_c$, so in order to compare them they must be given in the same dimensions.

Suppose cosmic rays are injected at $x=0$ when the background magnetic field $\mbfB = \hat xB$. We take $P_c(0,t)$ to be a given, increasing function of time and assume the
cosmic rays stream away from the boundary at speed $v_A$. To keep the problem simple, we assume $P_c(0,t) = P_{c0} + \Delta P_c(t)$ with $\Delta P_c/P_{c0}\ll 1$ and ignore the
compression and acceleration of the thermal gas by the cosmic rays. Then, $P_c$ at an interior point can be found from
\begin{equation}\label{eq:parallelstreaming}
\frac{\partial P_c}{\partial t} + v_A\frac{\partial P_c}{\partial x}=0,
\end{equation}
the solution of which is
\begin{equation}\label{eq:streamingsol}
P_c(x,t)=P_c(0,t-x/v_A),
\end{equation}
The heating rate $H_d$ is found directly from eqn. (\ref{eq:parallelstreaming})
\begin{equation}\label{eq:Hd}
H_d(x,t)=\frac{1}{v_A}\frac{\partial P_c(x,t)}{\partial t}=\Delta\dot P_c(0,t-x/v_A).
\end{equation}
Comparing eqns. (\ref{eq:hbound}) and (\ref{eq:Hd}), we see that heating due to pressure anisotropy is lower by a factor of order $v_A/c$, as suggested by the analysis in \S\ref{ss:Pressure}). The essential difference is that in the case of drift anisotropy, scattering
works against the tendency for cosmic rays to stream at the speed of light, but in the case of pressure anisotropy, scattering works against expansion at the magnetoacoustic speed.

\section{Implications for Cosmic Ray Self Confinement}\label{s:Implications}

Equations (\ref{eq:fpds}) and (\ref{eq:fppas}) constrain the product of the pitch angle scattering frequency $\nu$ and the anisotropy factor $A$, which appears in the growth rate $\Gamma_c$.   Requiring that $\Gamma_c$
 equals $\Gamma_d$, the rate of damping by  
 the thermal background, yields an independent constraint on $A$. From this, we can estimate the scattering frequency $\nu$, from which
we can derive the cosmic ray spatial diffusivity $\kappa\sim v^2/\nu$ and check for self consistency of the frequent scattering/short mean free path assumption that underlies cosmic ray hydrodynamics.

We can already guess from the results of \S\ref{ss:Comparison} that $\kappa_{pa}$, the diffusivity due to self confinement by pressure anisotropy instability, is larger than $\kappa_d$, its counterpart in the drift case, by a  factor of order $c/v_A$. From eqn. (\ref{eq:nu}) we see that $\nu$ is directly proportional to the fluctuation spectrum $W$, while  $\Gamma_c$ is now to be equated to $\Gamma_d$. If their product is smaller in the pressure anisotropy case by a factor of $v_A/c$, the scattering rates themselves must be smaller by approximately the same factor. Here, we provide some background on the argument and show explicitly how $\nu$ can be estimated. 

The most important damping mechanisms  are thought to be ion-neutral friction \citep{kulsrudpearce69}, nonlinear Landau damping by thermal ions \citep{leevolk73,kulsrud78}, damping by
an ambient turbulent  cascade \citep{yanlazarian02,farmergoldreich04}, and enhancement of turbulent damping by high plasma $\beta$ effects \citep{wieneretal2018}. With the exception
of ion-neutral friction, which appears to be strong enough to prevent cosmic ray self-confinement in dense, neutral gas entirely
\citep{everettzweibel11}, the other mechanisms suppress self confinement only above energies of about 100 - 300 GeV for Milky Way conditions. 
Although we
will not have to make explicit calculations involving any damping mechanisms to compare $\kappa_{pa}$ with $\kappa_d$, we provide a sample calculation for nonlinear Landau damping in \S\ref{s:Application}.

From eqns. (\ref{eq:fpds}) and (\ref{eq:Ad}) we estimate the diffusivity due to drift anisotropy instability as
\begin{equation}\label{eq:kappad}
\kappa_d\sim\frac{v^2}{\nu}\sim\left(v_AL\right)\left(\frac{v_D}{v_A}-1\right),
\end{equation}
which corresponds to a ratio of mean free path to lengthscale
\begin{equation}\label{eq:lambdad}
\frac{\lambda_d}{L}\sim\frac{v_A}{v}\left(\frac{v_D}{v_A}-1\right).
\end{equation}
If we take $\kappa_d$ to be the widely accepted value $3\times 10^{28}$ cm$^2$ s$^{-1}$, set $v_A=100$ km s$^{-1}$ (corresponding to an ion density of $0.01$ cm$^{-3}$ and $B=5\mu$G) and take
$L$ to be 1 kpc, then $v_D/v_A -1\sim 1$.

From eqns. (\ref{eq:fppas}) and (\ref{eq:Af2}), the diffusivity due to pressure anisotropy instability is
\begin{equation}\label{eq:kappapa}
\kappa_{pa}\sim v v_A\tau_B\left(\frac{c\Delta P_c}{v_AP_c}-1\right),
\end{equation}
where we have removed the spectrum dependent factors from eqn. (\ref{eq:Gammanpower}) because they are of order unity.
The corresponding ratio of mean free path to fiducial length $v_A\tau_B$
\begin{equation}\label{eq:lambdapa}
\frac{\lambda_{pa}}{v_A\tau_B}\sim \left(\frac{c\Delta P_c}{v_AP_c}-1\right).
\end{equation}
Due to the similarity between the drift and pressure anisotropy instability growth rates (eqns. \ref{eq:Gammacd} and \ref{eq:Gammanpower}), the drift and pressure anisotropy factors required to balance wave damping are probably about the same, namely, order unity. Therefore, the scattering mean free path due to pressure anisotropy is probably similar to the
Alfv\'en travel length $v_A\tau_B$.

\section{Example: Application to Bottlenecks in Galactic Halos}\label{s:Application}

One of the motivations for this study was the realization that models of cosmic ray self confinement by drift anisotropy should be prone to the formation of bottleneck. Here we apply
the theory of self confinement by pressure anisotropy to the models of bottleneck formation in  low density gas  between a cosmic ray source and a denser cloud such as shown in the top panel of Figure 1 \citep{wieneretal2017,wieneretal2019}. That study focused on the  effect of the cosmic ray source on the cloud. Here we address the effects of pressure anisotropy on the intercloud medium for the range of parameters chosen for that study. In \cite{wieneretal2019}, the cosmic rays were injected in a pulse of duration comparable to the Alfven
and sound travel times from the source to the cloud, so spatial gradients were important and a steady state was never achieved. In order to minimize these effects,we imagine that the cloud is much
closer to the source than the 1 kpc chosen in the original bottleneck studies.

Prior to cosmic ray injection, the intercloud medium has thermal pressure $P_g=3.2\times 10^{-13}$ dyne cm$^{-2}$, magnetic pressure $P_m= 3.97\times 10^{-14}$  dyne cm$^{-2}$,
and negligible cosmic ray pressure. A heating rate $\Gamma_0$ of $1.0\times 10^{-25}$ erg s$^{-1}$ per hydrogen atom is included to balance radiative losses at the initial temperature $T=1.1\times 10^6$  K, giving a volumetric heating rate of about $9.0\times 10^{-29}$ erg cm$^{-3}$ s$^{-1}$. 

The fiducial cosmic ray energy flux into the domain rate $\Delta\dot P_c$ is $1.67\times 10^{-25}$ erg cm$^{-3}$ s$^{-1}$. From eqn. (\ref{eq:dotn}), the characteristic flux tube transverse expansion time $\tau_B$ is $3.7\times 10^{12}$ s. Although the steady state cosmic ray pressure of $8.2\times 10^{-14}$ dyne cm$^{-2}$ derived from the simulation parameters is less than 25\% of the original
pressure, since $v_A/c\sim 2\times 10^{-4}$, the resulting distention of the magnetic field  is more than enough to  excite the cosmic ray  pressure anisotropy instability.

From eqn. (\ref{eq:hbound}), the lower bound on the heating rate $H_{pa}$ is $1.2\times 10^{-30}$ erg cm$^{-3}$ s$^{-1}$, slightly more than 1\% of the heating rate that offsets
radiative cooling. Even in models with ten times the fiducial source strength, heating would be a relatively weak effect compared to the heating required to offset radiative cooling..

The mean free path for scattering, however, is more interesting. From eqn. (\ref{eq:fppas}) We estimate $\nu$ by approximating eqn. (\ref{eq:fppas} as 
\begin{equation}\label{eq:approxnu}
\nu\frac{v_A}{c}p\frac{df_0}{dp}\left(\frac{\Delta P_c}{P_c}\frac{c}{v_A}-1\right)\sim\frac{2}{45}\frac{\dot B}{B}p\frac{df_0}{dp},
\end{equation}
from which it follows that
\begin{equation}\label{eq:nubottle}
\nu\left(\frac{c\Delta P_c}{v_AP_c}-1\right)\sim\frac{2c}{45v_A\tau_B}\sim 6\times 10^{-9}{\rm s}^{-1},
\end{equation}
corresponding to a mean free path of about 1.6 pc if the anisotropy factor is unity.  Bearing in mind that this value of $\nu$ is weighted by $\mu(1-\mu^2) < 1$, and that if wave damping
is weak the anisotropy factor could be significantly stronger, we conclude that pressure anisotropy instability could well be adequate to couple the cosmic rays to the intercloud
medium.

It's also instructive to repeat the estimate of $\nu$ if nonlinear Landau damping is the primary dissipation mechanism. For cosmic
rays near the mean energy, the damping rate $\Gamma_{nlld}$ is
\begin{equation}\label{eq:gammanlld}
\Gamma_{nlld}\sim\nu\frac{v_i}{c},
\end{equation}
where $v_i$ is the thermal ion velocity. Equating $\Gamma_{nlld}$  to $\Gamma_c$, which we estimate from eqn. (\ref{eq:fppas}),
gives
\begin{equation}\label{eq:nunl}
\nu\frac{v_i}{c}\sim\Gamma_c\sim\frac{\Omega_0}{\nu\tau_B}\frac{n_c}{n_i}\frac{c}{v_A}.
\end{equation}
Solving eqn. (\ref{eq:nunl}) for $\nu$, we find
\begin{equation}\label{eq:nusolution}
\nu\sim\left(\frac{\Omega_0}{\tau_B}\frac{n_c}{n_i}\frac{c^2}{v_iv_A}\right)^{1/2}.
\end{equation}
For the parameters assumed in \cite{wieneretal2019}, eqn. (\ref{eq:nusolution}) gives $\nu\sim 3.7\times 10^{-8}$s$^{-1}$. The
corresponding mean free path $\lambda\sim 0.26$ pc. Although this is a rough estimate, it suggests good confinement. Our estimate for the heating rate is unchanged. 

According to these results, waves driven by the pressure anisotropy instability would provide a short scattering mean free path and lock the cosmic rays to the thermal fluid. However,  the cosmic ray pressure profile would be quite flat and the cosmic rays would transfer little heat or momentum to the gas. 

\section{Summary and Conclusions}\label{s:Summary}

Cosmic ray propagation and cosmic ray hydrodynamics - the fluid description of how cosmic rays exchange energy and momentum with magnetized thermal plasma -  depend on
kinetic scale plasma processes. One of the most potent and best studied of these is 
 the drift, or streaming instability, of Alfven waves with a length scale of order the cosmic ray
gyroradius. In a steady state, which is assumed to be reached on a timescale short compared to the macroscopic dynamical time,  energy and momentum transferred  from the cosmic rays to the waves is absorbed by the thermal gas such that cosmic rays exert a parallel force of magnitude $\nabla_{\parallel}P_c$, heat the gas at a rate of magnitude $\vert v_A\nabla_{\parallel} P_c\vert$, and
drift at $\mathbf{v}_A$ relative to the thermal gas.

In order for the instability to operate, the parallel cosmic ray pressure and thermal gas density gradients must point in the same direction.  If this
 condition is not met, a cosmic ray ``bottleneck" forms, in which the cosmic ray pressure is constant and the cosmic rays do not exchange momentum or energy with the ambient medium. This can fundamentally alter the impact of cosmic rays on gas dynamics and thermodynamics, as speculated upon by \cite{Skilling71} and first demonstrated by \cite{wieneretal2017}.

In this paper, we have investigated a complementary mechanism for cosmic ray self confinement and coupling to the thermal gas based on an instability driven by cosmic ray pressure anisotropy. The pressure anisotropy instability, like the drift instability, is triggered by anisotropy of order $v_A/c$, destabilizes hydromagnetic waves with wavelength of order the cosmic ray gyroradius, and has a  growth rate proportional to the ratio of cosmic ray to thermal gas density, scaled by the nonrelativistic ion cyclotron frequency (eqns. \ref{eq:Gammacd} and \ref{eq:Gammanpower}).

In previous studies of this instability \citep{lazarianberesnyak06,yanlazarian11}, pressure anisotropy was assumed to be driven by compressive interstellar turbulence. In this situation, cosmic rays mediate the dissipation of turbulent energy
as heat, and absorb some of that energy themselves through second order Fermi acceleration. As such, the cosmic rays create an energy sink for the turbulent cascade at larger spatial scales
than would be expected due to dissipative processes in the thermal gas alone.

We considered anisotropy  driven by a slowly changing  magnetic field, but assumed  the energy source for  changing the  magnetic field to be the cosmic rays themselves, which we modeled as simple injection (\S\ref{s:formulation}; see Figure 1).  The anisotropy destabilizes circularly polarized waves, which propagate in both directions (\S\ref{s:instability}). In \S\ref{s:heating}, we calculated the relationship between the force exerted by the
cosmic rays on the medium and the rate at which they heat it. Whereas in the case of parallel streaming the magnitudes of the force and the heating rate are $\nabla_{\parallel}P_c$
and $v_A\nabla_{\parallel}P_c$, we found that for a transverse force $\nabla_{\perp}P_c$ the heating rate is only of order $(v_A/c)C_{ma}\nabla_{\perp}P_c$, where $C_{ma}$ is the magnetoacoustic speed defined in eqn. (\ref{eq:cma}). The underlying reason
for the difference in heating rates for drift and pressure anisotropy is that in the drift anisotropy case, scattering works against particle motion at the speed of light, but in the pressure anisotropy case in which expansion is transverse, the cosmic rays must overcome the inertia of the thermal gas and the driving is
at the much lower magnetoacoustic speed. We showed in \S\ref{s:Implications} that the weaker scattering rate corresponds to weaker self confinement.

In \S\ref{s:Application} we applied the pressure anisotropy model to one of the situations that motivated this paper: the formation of a bottleneck between an interstellar cloud and a cosmic ray source \citep{wieneretal2017,wieneretal2019}. We showed that although the heating rate is only a small perturbation to the thermodynamics, the scattering rate could be
enough to prevent the cosmic ray - thermal gas decoupling that would occur if a bottleneck formed. In this section we also showed how $\nu$ can be estimated when nonlinear damping is the main source of thermal dissipation; this led to an estimate for $\lambda$ somewhat shorter than the estimate based on linear damping.

Based on the calculations in this paper, we can say that pressure anisotropy instability and drift anisotropy instability are not equivalent and not interchangeable as far as cosmic ray confinement  and cosmic ray coupling to thermal gas on global (kpc)
scales are concerned. On intermediate, or mesoscales, the anisotropy drive may be strong enough to confine the cosmic rays and
provide momentum transport, but not a significant amount of heating. 

In general, cosmic ray sources are localized in space, and so in general, we would expect both drift and pressure anisotropy to be present.
For definiteness, suppose the direction of streaming is such that the drift anisotropy $f_1 > 0$ while $f_2 > 0$ due to cosmic ray
expansion of the flux tube onto which cosmic rays are injected. The waves for which drift and pressure anisotropy are both destabilizing have the largest growth rate. These  waves propagate in the same direction as the cosmic ray streaming and have $\mu_r > 0$. If $f_1$ and $f_2$ adjust such that cosmic ray excitation balances thermal damping, then waves propagating in the opposite direction, or with the opposite sense of
polarization, are damped. Therefore, the cosmic rays will be convected at the Alfv\'en speed, and their pressure will vary as $\rho_g^{\gamma_c/2}$ along the magnetic flux tube. If $\rho_g$ increases along the flux tube, this is exactly the condition for
 bottleneck formation. But due to pressure anisotropy, the cosmic rays do not decouple completely. Rather, as the pressure gradient
 flattens, counterpropagating waves of opposite polarization will become unstable, and the system will resemble the one studied
 here, with a short cosmic ray mean free path but little momentum or heat transfer. There would be very little difference between this bottleneck and the original one based on drift anisotropy alone. In future work, we hope to explore this complex picture  through analysis and numerical simulations.

\acknowledgements
I am happy to acknowledge useful comments by Chad Bustard, Adrian Fraser, Evan Heintz, Francisco Ley, Josh Wiener, and Huirong Yan and support by NSF Grant AST 1616037. Some of this work was accomplished at the Kavli Institute for Theoretical Physics at UC Santa Barbara.

\begin{appendix}

Here we sketch a proof that resonant scattering from an ensemble of randomly phased, small amplitude waves  transfers energy between waves and cosmic rays in a way that is consistent with wave growth and damping. For general discussions of wave/particle energetics see cite{kennel66}, cite{kulsrud69}, and cite{stix92}; here we consider only parallel propagating Alf\'ven waves, which is the relevant case for our problem.

Consider a wave of wavenumber $\mbfk=k\mbfB/B$ with electric field $\delta\mbfE_k$
\begin{equation}\label{eq:deltaEK}
\delta\mbfE(k)=\frac{1}{2}\left(\delta\mbfE_k e^{i\psi}+\delta\mbfE_k^{\ast}e^{-i\psi^{\ast}}\right),
\end{equation}
where $\psi\equiv kz-\omega t$. We will assume $\omega=\omega_r+i\Gamma_c$, with $\vert\Gamma_c/\omega_r\vert\ll 1$.
The spectral energy density
$\delta W_k$ is the sum of the electric, magnetic, and background plasma kinetic energy densities; the magnetic and kinetic energies are equal and larger
than the electric energy density by a factor of $(c/v_A)^2$. Then,
\begin{equation}\label{eq:Wk}
\delta W_k=2\frac{c^2}{v_A^2}\frac{1}{4\times 8\pi}\langle\left(\delta\mbfE e^{i\psi}+\delta\mbfE^{\ast}e^{-i\psi^{\ast}}\right)^2 \rangle=\frac{c^2}{v_A^2}\frac{\delta\mbfE_k\delta\mbfE_k^{\ast}}{8\pi},
\end{equation}
where  $\langle\rangle$ denotes an average over the phase $\psi$.

 Let the perturbed cosmic ray distribution function produced by the wave be $\delta f_{ck}$. In the quasilinear approach used here, diffusion in momentum space is produced
by the interaction of each wave electromagnetic field with the distribution function it produces, integrated over all $k$
\begin{equation}\label{eq:liouville}
\frac{Df_c}{Dt}= -\frac{1}{4}\int\langle q\left(\delta\mbfF_ke^{i\psi}+\delta\mbfF_k^{\ast}e^{-i\psi^{\ast}}\right)\cdot\frac{\partial}{\partial\mbfp}\left(\delta f_{ck}e^{i\psi}+\delta f_{ck}^{\ast}e^{-i\psi^{\ast}}\right)\rangle dk= -\frac{1}{4}\int\delta\mbfF_k\cdot\frac{\partial\delta f_{ck}^{\ast}}{\partial\mbfp}+\delta\mbfF_k^{\ast}\cdot\frac{\partial\delta f_{ck}}{\partial\mbfp}dk,
\end{equation}
where $\mbfF_k\equiv q(\delta\mbfE_k+\mbfvv\times\delta\mbfB_k/c)$ is the electromagnetic force due to the wave and $D/Dt$ is the convective derivative in phase space.
We derive an energy equation, by multiplying eqn. (\ref{eq:liouville}) by particle energy $\eps$ and integrating over momentum space. Here, we are mainly interested
in the right hand side, which we integrate by parts 
\begin{equation}\label{eq:rhs}
\begin{split}
\int\eps\frac{Df_c}{Dt}d^3p=-\frac{1}{4}\int\eps\delta\mbfF_k\cdot\frac{\partial\delta f_{ck}^{\ast}}{\partial\mbfp}+\delta\mbfF_k^{\ast}\cdot\frac{\partial\delta f_{ck}}{\partial\mbfp}dkd^3p =\frac{1}{4}\int q\mbfvv\cdot\left(\delta\mbfE_k\delta f_{ck}^{\ast}+\delta\mbfE_k^{\ast}\delta f_{ck}\right)dkd^3p  \\= \frac{1}{4}\int\left(\delta\mbfE_k\cdot\delta\mbfJ_{ck}^{\ast}+\delta\mbfE_k^{\ast}\cdot\delta\mbfJ_{ck}\right)dk,
\end{split}
\end{equation}
where $\delta\mbfJ_{ck}$ and $\delta\mbfJ_{ck}^{\ast}$ are the perturbed cosmic ray current Fourier component and its complex conjugate generated by the wave.

Each Fourier component of the cosmic ray current $\delta\mbfJ_{ck}$ is related to the total wave current $\delta\mbfJ_k$ and the wave plasma current
$\delta\mbfJ_{pk}$ by
\begin{equation}\label{eq:deltaJc1}
\delta\mbfJ_{ck} = \delta\mbfJ_k - \delta\mbfJ_{pk}.
\end{equation}
The relationship between $\delta\mbfE_k$ and $\delta\mbfJ_k$ is follows from combining Ampere's Law and Faraday's Law, and dropping the displacement current, which gives
\begin{equation}\label{eq:deltaJ}
\delta\mbfJ_k= \frac{ic^2k^2}{\omega}\frac{\delta\mbfE_k}{4\pi}.
\end{equation}
The plasma current $\delta\mbfJ_{kp}$ for an undamped, parallel propagating Alfven wave can be shown to be 
\begin{equation}\label{eq:deltaJp}
\delta\mbfJ_{pk}=\frac{i\omega c^2}{v_A^2}\frac{\delta\mbfE_k}{4\pi}.
\end{equation}
Substituting eqns. (\ref{eq:deltaJ}) and (\ref{eq:deltaJp}) into eqn. (\ref{eq:deltaJc1}) gives the cosmic ray current
\begin{equation}\label{eq:deltaJc2}
\delta\mbfJ_{ck}=i\frac{i\delta\mbfE_k}{4\pi}\frac{c^2}{v_A^2}\left(\omega -\frac{k^2v_A^2}{\omega}\right).
\end{equation}
Taking the scalar product of eqn. (\ref{eq:deltaJc2}) with $\delta\mbfE_k^{\ast}$, then taking the scalar product of the complex conjugate of eqn.
(\ref{eq:deltaJc2}) with   $\delta\mbfE_k$ and its complex conjugate, subtracting one equation from the other, and using eqn. (\ref{eq:Wk}) together
with the assumptions $\omega_r^2=k^2v_A^2$ and $\vert\Gamma_c/\omega_r\vert\ll 1$ gives
\begin{equation}\label{eq:jdotE}
\frac{1}{4}\left(\delta\mbfE_k\cdot\delta\mbfJ_{ck}^{\ast}+\delta\mbfE_k^{\ast}\cdot\delta\mbfJ_{ck}\right)=-2\Gamma_cW_k.
\end{equation}
Substituting eqn. (\ref{eq:jdotE}) into eqn.  (\ref{eq:rhs}) gives the result  we sought
\begin{equation}\label{eq:rhs2}
\int\eps\frac{Df_c}{Dt}d^3p=-\int 2\Gamma_{c}\delta W_kdk.
\end{equation}
Equation (\ref{eq:rhs2}) shows that the quasilinear force term represents the exchange of energy between the cosmic rays and the waves which scatter
them. If $\Gamma_c > 0$ (unstable waves), the cosmic rays lose energy to the waves.
\end{appendix}
\end{document}